\documentclass[traditabstract, bibyear]{aa} % for the abstract without structuration 
\usepackage{graphicx}
\usepackage{txfonts}
\usepackage{color}
\usepackage{amstext}
\usepackage{answers}

\begin{document}

   \title{Transit-period search from single-event space-based data: the role of wide-field surveys   
}

   \author{Geza Kovacs\inst{1}}

   \institute{Konkoly Observatory of the Hungarian Academy of Sciences, 
              Budapest, 1121 Konkoly Thege ut. 15-17, Hungary \\
              \email{kovacs@konkoly.hu ; kovageza@gmail.com}
             }

   \date{Received 23 February 2019 / Accepted 16 April 2019}

% \abstract{}{}{}{}{} 
% 5 {} token are mandatory
 
  \abstract
%{...}
%{...}
%{...}
%{...}
{We investigate the optimization of dataset weighting in searching for 
the orbital period of transiting planets when high-precision space-based 
data with a single transit event are combined with (relatively) 
low-precision ground-based (wide-field) data. The optimization stems 
from the lack of multiple events in the high-precision data and the 
likely presence of such events in the low-precision data. With noise 
minimization, we combined two types of frequency spectra: 
i) spectra that use two fixed transit parameters (moment of the center 
of the transit and duration of the event) derived from the space 
data alone; 
ii) spectra that result from the traditional weighted box signal search 
with optimized transit parameters for each trial period. We used many 
mock signals to test the detection power of the method. Marginal or no 
detections in the ground-based data may lead to secure detections in the 
combined data with the above weighting. Depending on the coverage and 
quality of the ground-based data, transit depths of $\sim0.05$\% and 
periods up to $\sim100$~days are accessible by the suggested optimum 
combination of the data. 
}

   
\titlerunning{Optimized transit-period search}
\authorrunning{Kovacs, G.}
   \maketitle
%
%________________________________________________________________

%
%%%%%%%%%%%%%%%%%%%%%%%
% SECTION 1
%%%%%%%%%%%%%%%%%%%%%%%
%
\section{Introduction}\label{introduction}
With the successful launch of the Transiting Exoplanet Survey Satellite (TESS) in 2018, it is conceivable 
that by the end of the two-year nominal mission we will have a complete 
census of hot Jupiters and Saturns (planets with radii larger than 
$\sim 0.3$~$R_{\rm J}$ and orbital periods shorter than $\sim10$~days). 

However, for most of the surveyed sky, systems with periods longer 
than $\sim 15$~days may only be marginally characterized because we 
lack reliable information on the orbital period. Exceptions are when 
at least two transits are observable due to a favorable observational 
window. For orbital periods longer than $\sim 30$~days, only single-transit 
events will be available (if at all) on $74$\% of the sky observed by 
TESS; see Yao et al.~\cite{yao2019}). According to Cooke et al.~(\cite{cook2018}),  
altogether, about $500$ planets will be observed only in single transits 
(see Huang et al.~\cite{huang2018} for a somewhat higher rate and 
Villanueva, Dragomir, Gaudi~\cite{villanueva2019} for another estimate). 
This represents some $5$--$10$\% of the planet population expected from 
the original mission (Huang et al.~\cite{huang2018}; Barclay, Pepper, 
Quintana~\cite{barclay2018}, and for a higher single-event rate 
Sullivan et al.~\cite{sullivan2015}). If the mission is to be extended, 
these targets will be reobserved, however, with a gap between the data 
acquired during the basic and the extended missions. The extent of the 
gap depends on the observational strategy to be followed in the extended 
mission (see Bouma et al.~\cite{bouma2017}). In spite of the gap, 
reobserving the same field may mean that the period question might be 
solved for systems that are also caught in transit during the second 
visit of the field. 

Because the transit shape depends on the orbital parameters, 
it might be possible to assess this important system parameter even in the 
case of single-transit event. Several studies have indeed been devoted 
to this topic (i.e., Yee \& Gaudi~\cite{yee2008}, Osborn et al.~\cite{osborn2016} 
and the earlier less extended work by Seager \& Mallen-Ornelas~\cite{seager2003} ;
for some deeper mathematical aspects, see also Kipping~\cite{kipping2018}). 
However, the probabilistic nature of this type of period estimate 
implies that the number of the photometric followup observations 
that is required to determine the period could require substantial resources 
even if the followup observations are planned carefully 
(Dzigan \& Zucker~\cite{dzigan2011, dzigan2013}). It is therefore 
important to search for other possible ways to constrain the period. 

If the estimated ratio of planet versus star mass is not too low, the period might be searched for by precise radial velocity observations because 
the variation is close to sinusoidal and therefore not time critical.  
Nevertheless, this method requires additional (usually expensive) spectroscopic 
telescope time and is therefore not always readily available.  

A third possibility is to search for signatures of the suspected signal 
in ground-based wide-field surveys, which have been running for at least 
several years (the two main surveys, the Super Wide Angle Search for Planets, 
SuperWASP, and the Hungarian-made Automated Telescope Network, HATNet, 
have run much longer; see Bakos et al.~\cite{bakos2004}, 
Pollacco et al.~\cite{pollacco2006}). Therefore, assuming that the 
transit is not too shallow, these surveys have a good chance to catch 
some events, even though they suffer from the quasi-daily gaps in the 
data sampling. Two recent works investigated this possibility more 
closely. Yao et al.~(\cite{yao2019}) examined the incidence rate of 
the discoverable transit signals in the Kilodegree Extremely Little 
Telescope (KELT) survey (Pepper et al.~\cite{pepper2007, pepper2012}) 
that are expected to appear as single transits in the TESS data. They 
derived discovery rates between $5$\% and $50$\% for systems hosting 
Saturn- to Jupiter-size planets, with periods of up to a year. 
Becker et al.~(\cite{becker2019}) used the data from HATNet, WASP, 
and KELT to confine the period of a long-period planet candidate 
in the multiplanetary system HIP~41378, discovered by 
Vanderburg et al.~(\cite{vanderburg2016}) from the data gathered 
by the Kepler two-wheel (K2) mission. They used the quality of the 
fit of the possible transit signals allowed by the K2 measurements 
to the ground-based data to attach probabilities to the various 
orbital periods. 

None of the above works analyzed space- and ground-based data in a 
combined fashion, in which one would work with a single statistics 
that would use the full dataset, and thereby reach maximum 
efficacy in detecting shallow signals. Here we introduce such a 
method, a part of which employs a search with fixed transit parameters 
derived from the high-precision single-transit space-based data 
(similar to the method used by Yao et al.~\cite{yao2019} in their 
detection survey). Throughout the paper we use mock data generated 
on the time base of real ground-based data and focus on the combination 
of these data with the single-sector TESS data (represented by pure 
mock data, including data sampling). We study the 
detection of shallow single transits with the aid of the combined 
data and assess the accessibility of the temperate to cold 
Neptune -- sub-Neptune regime. It is assumed that the data have already 
gone through the important step of systematics filtering, and the noise 
is white Gaussian. It follows that this study is for the investigation 
of the efficacy of the method presented, and not a population study, 
concerning actual detection rates for a given survey.  

%  
%%%%%%%%%%%%%%%%%%%%%%%
% SECTION 2
%%%%%%%%%%%%%%%%%%%%%%%
%
\section{Datasets, signal detection, and methods}
There are large number of possibilities for the actual observational 
settings of the wide-field ground- and space-based observations. 
This is mostly because of the wide ranges of the data distributions 
for the ground-based data, covering compact ($ \text{about one}$~month) and seasonal 
($ \text{about three to six}$~months) continuous (weather and day-time gaps limited) 
observations and those with multiple visitations over several years. 
There is also a range of noise properties, depending on the instrument 
used, site conditions, target brightness, etc. Obviously, it is 
impossible to assess the outcome of all possible settings. Instead, 
we select a few generic settings, and argue that the basic statements of 
this work on the optimization of the transit detection remain valid 
also in the non-generic cases. 

%  
%======================
% Subsection 2.1
%======================
%
\subsection{Time distribution}
\label{time_dist}
For the ground-based observations, the first data distribution we tested 
represents the ideal case that might become a reality when the large 
databases gathered by the various wide-field photometric surveys 
will be merged. For simplicity and to be more specific, we took one 
of the K2 targets that represents this ideal situation. Because of the 
high sampling rate (per instrument) of the ground-based survey data, 
we may consider the K2 time distribution taken as the $30$~min averages 
of the original data, under the ideal situation of continuous time 
coverage by the merged ground-based data. 

The second data distribution we tested are compact, seasonal datasets 
that are available through single-field observations from several sites, covering 
a certain range of longitude. We took the case of HAT-P-6 
(Noyes et al.~\cite{noyes2008}), which was observed between August and 
December 2005 by HATNet, which is a two-station network of small telescopes 
(Bakos et al.~\cite{bakos2004}). The data cover four months with almost 
ten thousand data points. The moments of time associated with these data 
were used to generate the test data. The effect of gaps between ground- 
and space-based data was tested by placing the ground-based data before 
the space-based data by an arbitrary amount of time, independently of 
the actual dates of the ground-based observations.  

The third dataset represents a reasonably common situation with 
multiple visitations of the same object over several years from 
several sites in overlapping fields. We took the case of HAT-P-2 
(Bakos et al.~\cite{bakos2007}), which was covered by HATNet, 
the Wise Hungarian-made Automated Telescope (WHAT), 
(Shporer et al.~\cite{shporer2009}) and SuperWASP 
(Pollacco et al.~\cite{pollacco2006}), comprising over 
forty thousand data points and covering $3.6$ years with gaps 
extending from some months to two years.

\begin{figure}
 \vspace{0pt}
 \centering
 \includegraphics[angle=-90,width=75mm]{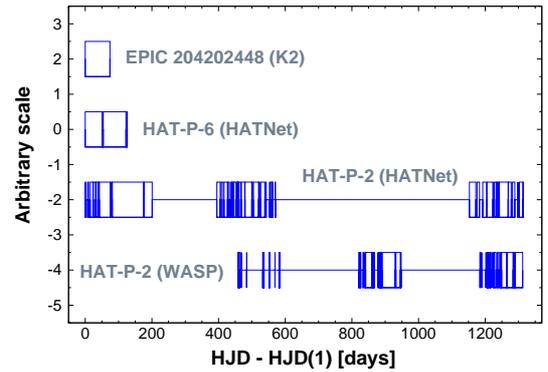}
 \caption{Time distribution of the secondary datasets used in 
         combination with the mock TESS data to test search methods for 
         periodic transit signals detected as single events in the TESS 
         data. Each block represents a continuous dataset with gaps shorter 
         than one day. The data are plotted relative to the time at the 
         first observation, HJD (1). The HATnet and WASP data on HAT-P-2 
         are merged in signal H2 (see Table~\ref{test_data}) for the 
         tests presented in this paper.}
\label{data_blocks}
\end{figure}

A symbolic representation of the time distributions of the ground-based 
data,  including the distribution of a representative K2 target to mock an idealized 
networked ground-based observation, is displayed in Fig.~\ref{data_blocks}.  
A brief summary of these data is given in Table~\ref{test_data}. The original 
HATNet and followup data are accessible through the respective publications 
and at the HATNet data site\footnote{\url{https://hatnet.org/}}. The WASP 
data on HAT-P-2 have been downloaded from one of the 
depositories\footnote{\url{https://wasp.cerit-sc.cz/form}} of the early 
data release of the project. 

To simulate the space-based data, we assumed a continuous cadence with 
half an hour sampling and $30$~days of total time coverage. These parameters 
are close\footnote{We rounded the nominal length of $27.4$~days of the 
single-sector coverage and omitted the relatively short gap of $16$ hours 
for pointing and data download at perigee (see Ricker et al.~\cite{ricker2015}, 
and to illustrate the quality of the TESS data, the full phase curve 
of WASP-18 by Shporer et al.~\cite{shporer2018})} to the characteristics 
of the single-sector data to be gathered by TESS after the completion of 
the two-year basic mission. Throughout this paper we refer to the 
space- and ground-based data as primary and secondary sets, respectively. 

%
%################
% Figure 1
%################
%
%

%
%
%================
% Table 1
%================
%
\begin{table}[!h]
%\centering
  \caption{Summary of the generic datasets.}
  \label{test_data}
  \scalebox{1.0}{
  \begin{tabular}{cccccc}
  \hline
  \hline
  \multicolumn{6}{c}{Datasets} \\
  \hline
   Set & Target & $N$  & $T_{\rm span}$[d] & $\Delta T$[yr] & Source \\ 
 \hline
H0  & EPIC-K2 &  $3585$ &   $74$ &  4 & {\small K2} \\
H1  & HAT-P-6 &  $9625$ &  $126$ & 13 & {\small HATNet} \\
H2  & HAT-P-2 & $45592$ & $1315$ & 10 & {\small H \& W} \\
T0  &    --   &  $1500$ &   $30$ &  0 & {\small MOCK} \\
\hline
\end{tabular}}
\\
  \scalebox{1.0}{
  \begin{tabular}{cccc}
  \multicolumn{4}{c}{Synthetic signals} \\
  \hline
  Type  & $P_{\rm orb}$[d] & $\delta$ & $\sigma_{\rm H}$ \\ 
  \hline
   a & $15.3-100$  & $0.001-0.003$  & $0.001-0.004$ \\
   b & $15.3-100$  & $0.001-0.003$  & $0.003-0.010$ \\
   c & $15.3-100$  & $0.003-0.010$  & $0.003-0.010$ \\
  \hline
\end{tabular}}
\begin{flushleft}
\underline{Notes:}
{\small 
From the data on the targets, only the time values of the respective 
time series are used in this paper (see Fig.~\ref{data_blocks}). H \& W denotes 
the merged HATNet and WASP data. $T_{\rm span}$ is the time span of the given 
dataset, $\Delta T$ is the difference between the last and first moments 
of time of the secondary (H\#) and the primary (T0) datasets, respectively. 
The time stamp on T0 is set to 2019.01.01. The transit depth $\delta$ is 
given in units of relative flux; $\sigma_{\rm H}$ is the point-by-point 
standard deviation of the white Gaussian noise added to the synthetic signals 
on sets H\#. The tranist duration is computed by assuming solar 
parameters for the host; see Winn~(\cite{winn2014}). The transit center 
is placed in the middle of T0. The standard deviation of the noise on the 
signals generated on T0 is set to be constant at $0.0005$ in all basic 
simulations. The average cadence is $30$~min for H0 and T0, and it is 
$5$~min for H1 and H2.}
\end{flushleft}
\end{table}
%

%  
%======================
% Subsection 2.2
%======================
%
\subsection{Transit signals}
\label{signals}
The basic parameters of the transit signals used in this work are 
given in Table~\ref{test_data}. We focused mostly on systems with 
Sun-like hosts and Neptune-like planets. To test single-transit 
events down to $\sim 15$~days, we fixed the center of the transit 
in the middle of the time span of the primary set T0. The three 
types of signal represent various signal--noise settings: 
type {\em a} for shallow signals in the presence of low noise, 
type {\em b} for shallow signals in the presence of medium to high 
noise, and type {\em c} for signals extending to the hot -Jupiter regime in the presence of medium to high noise. 

We chose $100$~days for the upper limit of the period to be searched 
for because this is about the extent of an observational season on a 
given field for the ground-based surveys. Furthermore, at an orbital 
period of $100$~days for a planet around a Sun-like star, we expect 
a transit duration of $0.36$~days, that is, far longer than the length of 
an average observation night. Although this difficulty can be 
overcome by even a modest longitudinal spread of a network of 
telescopes (e.g., HATNet), the problem of systematics filtering still 
remains for long events, comparable with the characteristic length 
of the continuous nightly data segments. In spite of all these difficulties, 
we note that ground-based surveys have already been successful in 
discovering systems with orbital periods longer than $10$~days. For instance, 
HATS-17b has an orbital period of $16.25$~days, with a transit duration 
of $0.20$~days (Brahm et al.~\cite{brahm2016}).  
The lower limit on the transit depths to be tested may seem overly 
optimistic, but we show that when combined with the single -event data from the primary (space) set, this limit is quite accessible 
(and even those down to $0.0005$ for more extended secondary sets). 
Ground-based surveys have already proven their ability to reach the 
few-millimagnitudes limit in detecting short-period transits (e.g., HAT-P-11 by 
Bakos et al.~\cite{bakos2010}, WASP-73 by Delrez et al.~\cite{delrez2014}). 
Because of this low limit in the detectable transit depths aided by 
ground-based data, the accessible systems cover a considerable upper 
region of the period--transit depth 
diagram\footnote{See \url{http://exoplanet.eu/} 
and \url{https://exoplanetarchive.ipac.caltech.edu/}}. We return to 
this aspect in Sect.~\ref{conclusions}. 

Except when indicated otherwise, for the primary dataset we fixed the 
standard deviation to $0.0005$ for the half-hour cadence. This 
value is more representative for the likely error budget of TESS at 
$\sim 11$~mag (i.e., Oelkers \& Stassun~\cite{oelkers2018}), whereas 
for brighter stars, $0.0002$ or even lower values may be used 
(e.g., Shporer et al.~\cite{shporer2018}). From our point of view, 
the exact value does not really matter until it is considerably 
(i.e., by several factors) lower than the standard deviation of the 
secondary dataset. For the latter, the error ranges are realistic 
in the bright tail of the magnitude distribution (e.g., between $9$ 
and $11$~mag) for all signal types listed in Table~\ref{test_data}; see Bakos et al.~(\cite{bakos2009}). For signal type~{\em a,} the 
errors are realistic only if the data are averaged on a $30$~min 
cadence (i.e., for set H0). 

In most of our tests we used $500$ random realizations for each test case. 
In these realizations we chose the transit parameter $\delta$ and the 
standard deviation $\sigma_{\rm H}$ of the white-noise component of the 
signal from a uniform distribution. The same type of distribution was 
used for the orbital period, but this time, for a better sampling of 
the shorter period regime, on the logarithmic values of the period. 
The transit length was computed for each test period by assuming a central 
transit, a circular orbit, and a solar-type host (e.g., Winn~\cite{winn2014}). 
The distribution of the noise component was Gaussian, with the standard 
deviation chosen above. We stress that these random simulations are not 
intended to model the observed extrasolar planet population. Instead, our 
purpose is solely to visit a wide but still plausible parameter space.  

\begin{figure}
 \vspace{0pt}
 \centering
 \includegraphics[angle=-90,width=75mm]{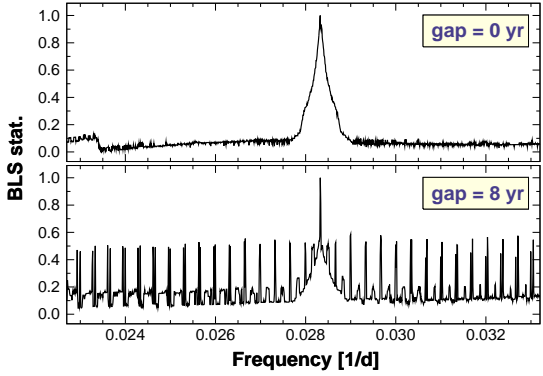}
 \caption{Comparison of the BLS spectra obtained without (upper 
        panel) or with a single $8$~yr gap between the ground-based and 
        TESS data. We use the datasets H0 and T0 (emulating the TESS 
        data) of Table~\ref{test_data} with a noiseless signal of 
        $P=35.3$~days and a relative transit duration $t14/P=0.008$. 
        Each spectrum is normalized to the respective highest 
	peak.}
\label{gapped_bls}
\end{figure}

Last but not least, as listed in Table~\ref{test_data}, in the 
actual joint analysis with the already existing ground-based data, we 
expect rather large gaps between these and the TESS data. This is 
a serious problem both from the point of the practical implementation 
of the signal search (the required number of test frequencies might 
easily exceed $10^5$--$10^6$ over the frequency band of $\sim 0.1$~d$^{-1}$) 
and it is also bad for the signal-to-noise ratio (S/N) and thereby 
for the reliability of the signal search. To show the effect of a single 
gap of $8$~yr between sets H0 and T0, we compare in Fig.~\ref{gapped_bls} 
the resulting box-fitting least-squares (BLS) spectrum 
(Kovacs et al.~\cite{kovacs2002}) with the spectrum obtained in the case 
of continuous data distribution. Although the main peak preserves the 
overall width of the line profile, there are subtle differences, in addition 
to the appearance of a forest of peaks at a lower power. When combined with 
noise, these aliasing effects lead to lower S/Ns, and ultimately to a lower 
discovery rate for data that are separated by longer gaps. Because of the 
practical importance of this effect, we also tested how gaps influence 
the results that are obtained by the optimized joint analysis presented 
in this paper. 

%
%################
% Figure 2
%################
%
%

%  
%======================
% Subsection 2.3
%======================
%
\subsection{BLS spectrum characterization and detection criteria}
\label{bls_params}
Because the method presented here is based on the optimization of the 
BLS frequency spectra, we here briefly summarize some of the technical 
details in the computation of the spectra and the parameters used 
to characterize the detection and signal significances. On the 
basis of these parameters we also define the detection criteria to 
be used throughout the paper to classify the various signal search 
methods.

%
%################
% Figure 3
%################
%
\begin{figure}
 \vspace{0pt}
 \centering
 \includegraphics[angle=-90,width=75mm]{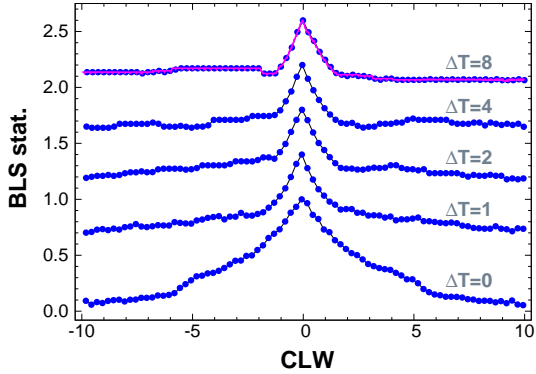}
 \caption{Evolution of the line profile of the BLS frequency spectrum 
        as a function of the gap length ($\Delta T$~[yr]) between 
        two sets of time series (H0 and T0 of Table~\ref{test_data}). 
        The same signal is used as in Fig.~\ref{gapped_bls}. 
        Four samples for each CLW (see text) are used (blue 
        points). For comparison, the densely sampled profile is also 
        shown by the pink continuous line. All profiles are normalized 
        by the peak power and shifted vertically for better visibility. 
        The apparent lift of the profile wings for higher $\Delta T$ 
        values is due to the CLW units used on the abscissa.}    
\label{profile_sampling}
\end{figure}

For the basic tests, all frequency spectra were computed in the 
$[\nu_{\rm min},0.15]$~d$^{-1}$ range, where $\nu_{\rm min}=1/T_{\rm span}$, 
and $T_{\rm span}$ is the total time span of the analyzed 
data.\footnote{We employed a lower limit in the frequency search to 
avoid large gaps in the folded time series when we computed the BLS 
statistics.} For frequency sampling, in the course of extensive parameter 
survey, we used four samples for each central line width CLW~$\sim q_{\rm tran}/T_{\rm span}$, 
where $q_{\rm tran}=t14/P_{\rm orb}$ is the relative transit length 
(the ratio of the full transit duration $t14$ to the orbital period 
$P_{\rm orb}$). This sampling has a coverage of $ \text{about eight}$ 
spectral points throughout the full central part of the line profile 
(see Fig.~\ref{profile_sampling}). 

The S/N of the frequency spectrum, $\rm S/N_{\rm sp}$ , always 
referred to a given frequency band,  is naturally defined as 
%
%**************
% Eq. (1)
%**************
%
\begin{eqnarray}
\label{snr_sp_def}
\rm S/N_{\rm sp} = {SP(\nu_{\rm peak}) - <SP> \over \sigma(SP)} \hskip 1mm ,
\end{eqnarray}  
where $SP(\nu_{\rm peak})$ is the value of the BLS power at the peak 
frequency $\nu_{\rm peak}$, $<SP>$ is the average power, and $\sigma(SP)$ 
is the standard deviation in the given frequency band. These quantities 
are derived from the BLS spectra after subtracting a best-fitting 
$\text{six}$th-order polynomial from the original spectra and normalizing 
it to $[min,max]=[0,1]$. The polynomial fit is necessary to eliminate the 
common overall power increase at low frequency in the BLS spectra. 
Therefore, we employed an iterative fitting that discards outliers, 
that is, high peaks, at the $3\sigma$ level.  

It might be a matter of dispute how to derive $<SP>$ and $\sigma(SP)$ 
because in the case of gapped data, aliasing produces additional peaks 
that might increase both the average and the standard deviation of the 
spectrum. However, periodic signals have asymptotically discrete spectra 
also when the data are gapped, and the straightforward computation of 
$<SP>$ and $\sigma(SP)$ could also be sufficient because the peaks still 
occupy only a small fraction of the frequency band investigated if this 
band is wide enough. Therefore, we decided not to use outlier clipping 
when we computed $<SP>$ and $\sigma(SP)$. 

In addition to the spectrum S/N, we might be interested also in the 
significance of the signal in the folded light curve (LC) and ask 
whether any conclusion might be derived from the quality of the folded 
LC of some hypothetical signal on the detectability of this signal in 
the BLS spectrum. Following Kovacs \& Kovacs~(\cite{kovacs2019}), we 
considered the immediate neighborhood of the transit with the same 
length of out-of-transit section as the transit itself. When we assume 
a uniform data distribution with $N_{\rm in}$ intransit, $\delta$ transit 
depth and point-by-point errors with a standard deviation $\sigma$, 
the significance of the transit is parameterized simply by the ratio 
of the transit depth and the error of the difference between the 
averages of the in- and out-of-transit parts, i.e.,  
%
%**************
% Eq. (2)
%**************
%
\begin{eqnarray}
\label{snr_lc_def}
{\rm S/N}_{\rm lc} & = & {\sqrt{N_{\rm in} \over 2}} {\delta \over \sigma} \hskip 1mm .
\end{eqnarray}  

To quantify the power of the various detection methods, we need to define 
the criteria of detection. Since we investigate test signals with known 
parameters, we can easily define these conditions as follows: 
\begin{itemize}
\item[a)]
The S/N of the highest peak in the frequency spectrum, 
$\rm S/N_{\rm sp}$, must be greater than $\rm S/N_{\rm min}$, the 
lower detection limit, set according to the acceptable tolerance 
on the false-alarm rate (FAR): the rate of those spectra that satisfy the condition on $\rm S/N_{\rm sp}$, 
but do not satisfy the frequency condition below. 
\item[b)]
The frequency at the highest peak, $\nu_{\rm peak}$, should satisfy the 
following condition: $|\nu_{\rm peak}-r\nu_{\rm test}| < 2\times{\rm CLW}$, 
where $r=n/m$, with small integers to allow traceable frequency confusion, 
and CLW$\sim q_{\rm tran}/T_{\rm span}$, as already defined earlier. 
\end{itemize}
Including the test frequency, we check altogether $15$ frequencies of the 
type above (i.e., we do not include alias components due to sampling). 
Interestingly, we find that in the large majority of cases, it is the basic 
frequency that comes out as the largest peak in the spectra. In the case 
of real data, of course, we do not know FAR for any given $\rm S/N_{\rm min}$. 
Usually, as a rule of thumb, by taking $\rm S/N_{\rm min}\sim 6-8$ will result 
`meaningful' FAR values, i.e., less than $\sim 20$\%. To get a more accurate 
estimate on FAR when real data are analyzed, one can perform an injected 
signal test. This will certainly increase the execution time, but supplies 
an important piece of information on the reliability of the suspected 
detection. Furthermore, a deeper examination of the spectra (e.g., alias 
search, including all peaks, not only the highest one) would certainly 
increase the detection rate by some -- foreseeably small -- amount.  
We opted not to dwell so deeply in the analysis of the frequency spectra, 
because the expected gain is small, and we are interested in relative 
detection rates, and then employing the same detection method is more 
important than getting a little gain by a deeper spectrum analysis. 
We return to the issue of detection rate and detection thresholds in 
Sect.~\ref{efficacy}.      

%  
%======================
% Subsection 2.4
%======================
%
\subsection{Optimizing $\rm S/N_{\rm sp}$ for the joint BLS spectrum}
\label{bls3}
With the far better quality of the TESS data (represented by set T0 in 
Table~\ref{test_data}), it is obvious that traditional inverse variance 
weighting in the least-squares transit search of the joint data would not 
work with single-event TESS data. Therefore, we propose to search for an 
optimum weighting with the aid of simple scanning a range of weights, and 
search for the weight that yields the highest S/N for 
the corresponding BLS frequency spectrum. In the following we describe 
the basic ingredients on which our detection analysis is based, and present 
some examples exhibiting the characteristics of these ingredients.  

%
%################
% Figure 4
%################
%
\begin{figure}
 \vspace{0pt}
 \centering
 \includegraphics[angle=0,width=75mm]{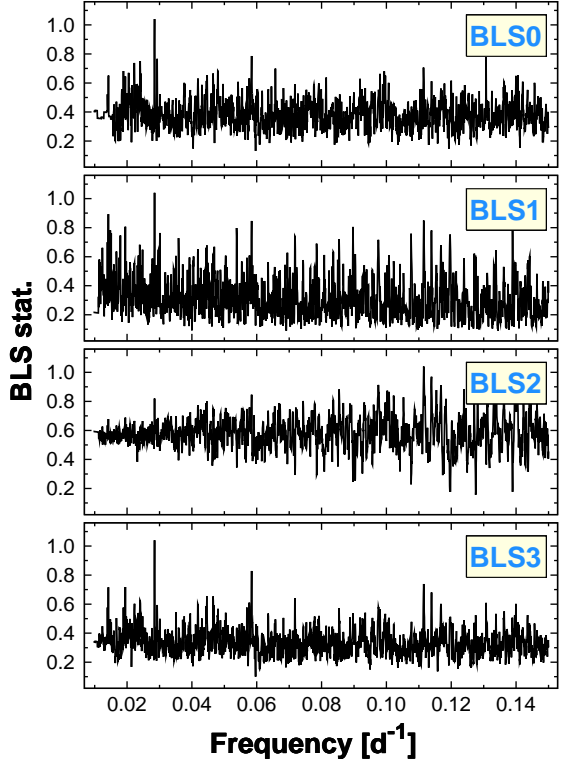}
 \caption{Typical spectra obtained by variants of BLS applied on set 
        H0+T0 of Table~\ref{test_data}. The transit parameters are  $P_{\rm orb}=35.12345$~d, $\delta=0.003$, $q=0.005$, 
        $\sigma_{H0}=0.002$, and $\sigma_{T0}=0.0005$. There is no gap between 
        H0 and T0. Economic plotting is employed using $2000$ frequency 
        bins and plotting only the maxima in each bin. Increasing 
        the number of fixed transit parameters (from the `all free' case 
        of BLS0 to the 'all fixed' case of BLS2) decreases the detection 
        efficacy in this faint signal regime. However, the proposed 
        optimum combination of BLS0 and BLS1 to BLS3 significantly increases 
        the reliability of the detection.}
\label{bls0123}
\end{figure}

First, we introduce a single parameter $\alpha$ that was used to weight 
set T0. For simplicity, we used equal weights on the data associated 
with the same source, that is, use $\alpha$ for set T0 and $1-\alpha$ 
for sets H\#. These weights can be employed directly in the formulae 
of Kovacs et al.~(\cite{kovacs2002}) to compute the BLS spectra because 
these formulae allow arbitrary weighting of the data. Then we may 
consider various possibilities to perform the BLS analysis. In a 
basic setting we ignore the fact that T0 consists of data of 
considerably higher accuracy than H\#, and with a properly chosen 
$\alpha,$ we performed a traditional BLS analysis by letting all transit 
parameters free to vary (but least-squares optimized) at each test 
period. We call this approach BLS0. 

Then, following in part Yao et al.~(\cite{yao2019}), we derived the 
time of the transit center $T_{\rm c}$ and duration $t14$ from T0, 
and then scanned only the period to find the one that fits the full 
dataset. The scanning was performed within the framework of weighted 
BLS, whereby the transit depth is optimized at each trial period. 
We flagged the resulting spectrum as BLS1. An alternative approach 
would be try to fix the transit depth and scan only the 
period.\footnote{This was also mentioned in Yao et al.~(\cite{yao2019}), 
but was discarded as an improper approach because of the complications 
that might arise from the different quality of the KELT and TESS images. 
We also discarded this method, but for a different reason 
(see Sect.~\ref{bls2_stat}).} This approach is labeled BLS2. 
  
Before proceeding to the fourth method, we show an example of the 
methods introduced so far in Fig.~\ref{bls0123}. The standard 
BLS method with weights (uppermost panel) is able to detect the signal 
with a reasonable significance. This can be compared with BLS1, where 
the signal is apparently less significant. However, a deeper 
examination of the plots reveal that the noise baseline (average noise 
level) is higher for BLS0. Because this parameter also plays role in 
the calculation of $\rm S/N_{\rm sp}$ (see Eq.~\ref{snr_sp_def}), the 
situation is more complex than it may seem at first sight. 
Nevertheless, it is also quite clear that the two types of spectra 
display relatively little correlation in the noise-dominated frequency 
regime. This prompted us to attempt to improve the method and 
take the weighted average of the two spectra, leading to BLS3, with the 
corresponding spectra defined as 
%
%**************
% Eq. (3)
%**************
%
\begin{eqnarray}
\label{sp3_def}
SP_3(\nu) = \beta\times SP_0(\nu) + (1-\beta)\times SP_1(\nu) \hskip 1mm , \hskip 4mm 
\beta = {s_1^2 \over s_0^2 + s_1^2} \hskip 1mm , 
\end{eqnarray}  
where $SP_{0,1,3}(\nu)$ and $s_{0,1}$ are the spectra and the standard 
deviations, respectively, corresponding to the methods described above 
(as indicated by the subscripts). We note that the spectra are normalized 
to [min,max]=[0,1] and $s_{0,1}$ refer to these spectra. The BLS1 spectrum 
is optimized separately. The weight for BLS1 is always close to a low value, 
and the $\rm S/N_{\rm sp}$ dependence on $\alpha$ is rather weak (see below), 
therefore we fixed $\alpha$ for BLS1. The bottom panel of Fig.~\ref{bls0123} 
clearly shows the positive effect of the averaging and indicates that this 
method is probably preferred over the other alternatives discussed 
above for joint signal analysis. 

%
%================
% Table 2
%================
%
\begin{table}[!h]
%\centering
  \caption{Summary of the variants of the BLS routine.}
  \label{bls_def}
  \scalebox{1.0}{
  \begin{tabular}{lll}
  \hline
  \hline
   Name & Fixed parameters & Description \\ 
 \hline
BLS0  & None                    & BLS spectrum S/N maximization,\\
      &                         & uniform weights:\\ 
      &                         & T0 [$\alpha$], H\# [$1-\alpha$]\\
BLS1  & $T_{\rm c}$, $t14$            & zero point and $\delta$ fit,\\
      &                         & optimized weights as for BLS0\\
BLS2  & $T_{\rm c}$, $t14$, $\delta$  & zero point fit,\\
      &                         & optimized weights as for BLS0\\
BLS3  & BLS0$+$BLS1             & BLS spectra are inverse variance\\
      &                         & weighted (see Eq.~\ref{sp3_def}); \\
      &                         & data weights ($\alpha$, see text) are kept\\
      &                         & fixed for BLS1 but varied for\\ 
      &                         & BLS0 to yield the maximum\\ 
      &                         & S/N for the BLS3 spectrum\\
\hline
\end{tabular}}
\end{table}

We return to BLS2. It is interesting to realize that fixing all transit 
parameters except for the period leads to a very high sensitivity to 
noise and disqualifies BLS2 as a powerful method in detecting faint 
signals. We note that we rejected BLS2 for 
generic reasons, whereas Yao et al.~(\cite{yao2019}) rejected it based on the possible 
difficulties of combining different data from telescopes of considerably 
different optical properties. It is also observable that although the 
signal is detected with a low significance in the immediate neighborhood 
of the true frequency, this is spoiled by the increasing noise at higher 
frequencies. To understand this unexpected behavior, we examine the 
statistical properties of the BLS2 spectra under certain idealized conditions 
in Sect.~\ref{bls2_stat}. For easier reference, we briefly summarize the main ingredients of 
each method in Table~\ref{bls_def}. 

\begin{figure}
 \vspace{0pt}
 \centering
 \includegraphics[angle=-90,width=75mm]{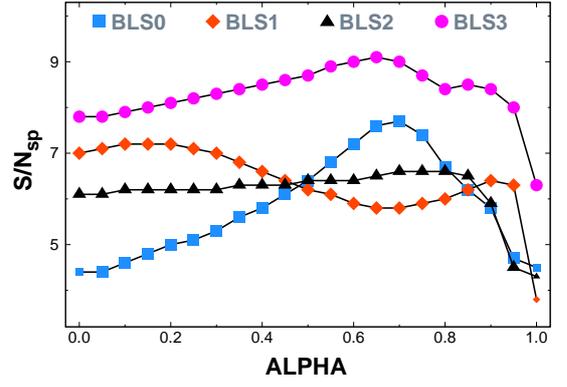}
 \caption{Signal-to-noise ratio of the highest peak of the BLS spectra 
          vs data weight factor $\alpha$ for a realization of a faint 
          transit signal on the set H0+T0 with zero gap. The transit 
          parameters are as follows: $P_{\rm orb}=15.2345$~d, 
          $\delta=0.003$, $q=0.010$, $\sigma_{\rm H0}=0.004$, 
          $\sigma_{\rm T0}=0.0005$. All peak frequencies match the orbital 
          frequency, except the very few shown by smaller symbols. See 
          Table~\ref{bls_def} and associated text for the definition 
          of the various BLS methods.} 
\label{alpha_scan3_low}
\end{figure}

\begin{figure}
 \vspace{0pt}
 \centering
 \includegraphics[angle=-90,width=75mm]{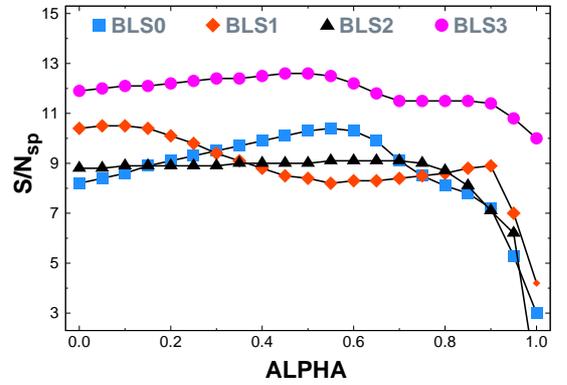}
 \caption{As in Fig.~\ref{alpha_scan3_low}, but for a stronger signal 
 where all parameters are the same, except for the increased transit 
 depth of $\delta=0.005$. The increased signal strength results in a 
 shift of the optimum $\alpha$ toward lower values and some flattening 
 of the $\alpha$ dependence of $\rm S/N_{\rm sp}$ for BLS0 (and, as a 
 result, also for BLS3).}
\label{alpha_scan3_high}
\end{figure}

\begin{figure}
 \vspace{0pt}
 \centering
 \includegraphics[angle=-90,width=75mm]{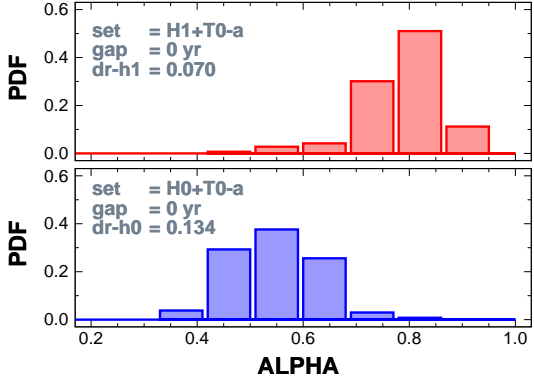}
 \caption{Distribution of the optimum $\alpha$ values from the 
          BLS0 analyses of datasets of different signal detection 
          powers for the secondary sets (see Table~\ref{test_data}). 
          The detection ratios obtained from the BLS analyses of 
          the secondary sets alone are given by dr-h0 and dr-h1. 
          All data points used in the construction of the histogram 
          satisfy the detection criteria with $\rm S/N_{\rm min}=6$. 
          We use the same $500$ realizations in both datasets. 
          Aliasing in set H1 boosts $\alpha$ to higher values for 
          optimized BLS0 performance.}
\label{two_alpha}
\end{figure}

For the maximization of $\rm S/N_{\rm sp}$ for BLS0 and BLS1, it 
is important to examine whether any prediction can be made about 
the associated weights from the bulk statistical properties of the 
constituting time series. Based on a specific time series, 
Figs.~\ref{alpha_scan3_low} and \ref{alpha_scan3_high} show examples 
of the behavior observed in most of the tests we performed.  
The two figures are meant to illustrate the different dependence of 
$\rm S/N_{\rm sp}$ on $\alpha$ in the low- and high-amplitude regimes  
(shown in Figs.~\ref{alpha_scan3_low} and \ref{alpha_scan3_high}, respectively). The following general properties emerge 
from inspecting these two figures: 
\begin{itemize} 
\item[-]
$\rm S/N_{\rm sp}$ has a strong dependence on $\alpha$ for BLS0, especially 
for weaker signals. 
\item[-]
The maximum of $\rm S/N_{\rm sp}$ for BLS0 shifts to lower $\alpha$ for 
stronger signals. 
\item[-]
BLS1 has nearly flat maximum $\rm S/N_{\rm sp}$ in the $[0,0.2]$ regime, 
independently of the strength of the signal. 
\item[-]
BLS2 has nearly flat maximum $\rm S/N_{\rm sp}$ in a very wide regime 
in $[0,0.8]$. 
\item[-]
$\rm S/N_{\rm sp}$ for BLS3 is flatter than BLS0, but has very similar 
properties.  
\item[-]
BLS3 has the highest maximum $\rm S/N_{\rm sp}$ among the tested methods. 
\end{itemize} 
Of the listed properties, the overall shift of $\alpha$ toward 
lower values for higher-quality secondary datasets seems fairly 
robust because it is also detected in the more extensive tests made on  
datasets H1+T0 and H0+T0. Figure~\ref{two_alpha} shows the resulting 
distributions for $\alpha$ from the $500$ simulations for both 
data settings, using the same realizations (which means that the only difference 
between the two cases is the data distribution: H0 is continuous, H1 is 
gapped, but contains more data points). Because of the aliasing, the 
detection rate in H1 alone (dr-h1) is lower than in set H0. Therefore
the effect on the optimum $\alpha$ is similar to the case of a lower 
signal amplitude, as shown in Figs.~\ref{alpha_scan3_low} and 
\ref{alpha_scan3_high}. 

Except for the properties above, we were unable to derive any simple rule 
to determine $\alpha$ for BLS0/BLS3 and avoid the somewhat time-consuming 
search for the optimum value of this parameter. However, for BLS1 we can 
safely fix $\alpha$ at $\sim 0.1$ for a wide regime of data and signal 
parameters. 

%
%################
% Figure 5
%################
%
%

%
%################
% Figure 6
%################
%
%

%
%################
% Figure 7
%################
%
%

%  
%======================
% Subsection 2.5
%======================
%
\subsection{Statistical properties of BLS2}
\label{bls2_stat}
We assumed that 
(i) the secondary dataset contains only a single transit, and 
(ii) only those trial periods are tested that generate transits 
that either have full overlap or no overlap at all  
with the true transit in the secondary dataset. The time 
series of the secondary set has the form of $x(i)=T(i)+\xi(i)$, 
where $T(i)=\delta$, if $i_1 < i < i_2$, and zero for all the 
other $n$ data points of $\{x\}$. The noise component $\{\xi\}$ 
is assumed to be Gaussian and white, i.e., the following relations 
are held for the expectation values: $<\xi>=0$, $<\xi^2>=\sigma^2$, 
$<\xi^3>=0$, $<\xi^4>=3\sigma^4$ and $<\xi(i)\xi(j)>=0$ for any $i\neq j$. 
When the correct signal frequency $\nu_0$ is hit, the average squared 
deviation of the residuals is computed by   
%
%**************
% Eq. 4
%**************
%
\begin{eqnarray}
\label{eq_4}
D(\nu_0) & = & {1 \over n} \sum_{i=1}^{n} \xi_i^2 \hskip 1mm . 
\end{eqnarray}  
For test periods that do not yield overlap with the single event 
in the secondary dataset, there are $k$ trial transits in the out-of-transit 
part of the data, and there is an out-of-transit part of the trial signal 
at the transit section of the data. Assuming that all transits include 
the same number of $m$ data points, we have 
%
%**************
% Eq. 5
%**************
%
\begin{eqnarray}
\label{eq_5}
D(\nu) & = & {1 \over n} (\sum_{i=1}^{n-m(k+1)} \xi_i^2 + \sum_{j=1}^{m(k+1)} (\delta-\eta_j)^2) \hskip 1mm ,  
\end{eqnarray}  
where $\{\eta\}$ is some subset of the full set of $\{\xi\}$ and disjunct 
from the set entering in the first sum. The first two moments of $D(\nu_0)$ 
and $D(\nu)$ can easily be derived from the above expressions and the statistical 
properties of $\{\xi\}$. We find   
%
%**************
% Eq. 6
%**************
%
\begin{eqnarray}
\label{eq_6}
<D(\nu_0)> & = & \sigma^2 \hskip 2mm ; \hskip 2mm \sigma^2(D(\nu_0)) = {2\sigma^4\over n} \hskip 2mm ,   
\end{eqnarray}  
%
%
%**************
% Eq. 7
%**************
%
\begin{eqnarray}
\label{eq_7}
<D(\nu)>         & = & \sigma^2 + {m(k+1) \over n}\delta^2 \hskip 2mm , \nonumber \\
\sigma^2(D(\nu)) & = & {2\sigma^4\over n} + {4m(k+1) \over n^2}\delta^2\sigma^2 \hskip 2mm ,    
\end{eqnarray}  
where $\sigma^2(D(\nu))=<D(\nu)^2> - <D(\nu)>^2$. As expected,  
the spectra show an overall linear increase in the average power toward 
higher frequencies. In principle, this would not be a problem because the 
spectra are filtered out from polynomial trends. However, the similar 
increase in frequency-dependent variance degrades the 
result. When we normalize the variance of the spectrum 
to the value at the true frequency, we obtain for the relative increase of 
the variance $Q_k=2(k+1)(m/n)(\delta/\sigma)^2$, which confirmes what 
we see in the actual numerical tests (see Fig.~\ref{bls0123}).

%  
%%%%%%%%%%%%%%%%%%%%%%%
% SECTION 3
%%%%%%%%%%%%%%%%%%%%%%%
%
\section{Efficacy of the joint analysis}
\label{efficacy}
Before we investigate the power of the joint analysis, we briefly discuss 
the basic patterns of the detection rates and FARs that are 
the basic parameters for comparing the power of the various detection 
methods. We define two types of detection rate: (i) the observed rate 
$\rm DR_{\rm obs}$, where the only requirement is to satisfy the 
$\rm S/N_{\rm sp}^{\rm peak} > \rm S/N_{\rm sp}^{\rm min}$ criterion, and 
(ii) the true rate $\rm DR_{\rm true}$, satisfying both the above and the 
frequency match criteria (see Sect.~\ref{bls_params}). We denote the 
number of cases that satisfy condition (i) by $N_{\rm S/N}$, and those 
that also satisfy the frequency condition by $N_{\rm S/N, \nu}$. We recall 
that the FAR is then simply $FAR=1.0-N_{\rm S/N, \nu}/N_{\rm S/N}$. 
It follows then that $\rm DR_{\rm true}=(1-{\rm FAR)}\times DR_{\rm obs}$.   

%
%################
% Figure 8
%################
%
\begin{figure}
 \vspace{0pt}
 \centering
 \includegraphics[angle=-90,width=75mm]{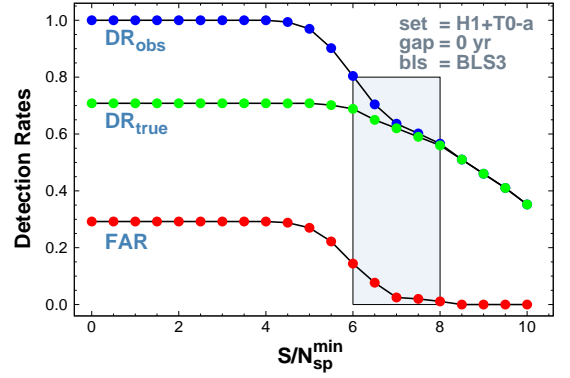}
 \caption{Dependence of the detection rate on the lower limit of the 
          spectrum S/N. The observed rates 
          $\rm DR_{\rm obs}$ are calculated solely on the basis of satisfying 
          the $\rm S/N > \rm S/N_{\rm min}$ criterion and do not consider the 
          match of the peak frequency to the test frequency. The true 
          detection rate $\rm DR_{\rm true}$ results from the correction 
          of the observed rate by the FAR, namely, 
          $\rm DR_{\rm true}=(1-{\rm FAR)}\times DR_{\rm obs}$. The signal 
          identification and analysis method are indicated in the 
          upper right corner. The most commonly used range for the lower 
	  limit of the spectrum S/N is shown by the gray rectangle. 
	  We use $500$ random realizations of the signal shown in the 
	  header; see also Table~\ref{test_data}.}
\label{snr_vs_fap_0}
\end{figure}
%

%
%################
% Figure 9
%################
%
\begin{figure}
 \vspace{0pt}
 \centering
 \includegraphics[angle=-90,width=75mm]{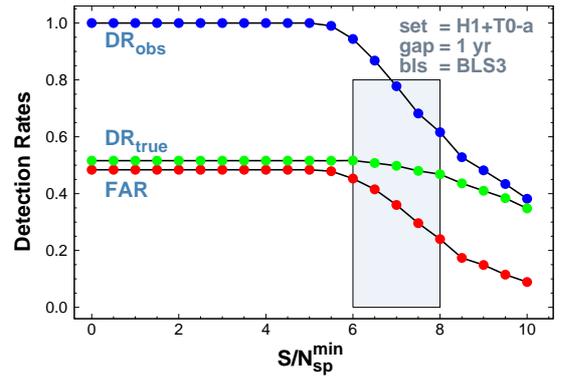}
 \caption{As in Fig.~\ref{snr_vs_fap_0}, but for the simulations 
          with $1$~yr gap between the primary and secondary datasets. 
          Aliasing leads to higher FARs and concomitant 
          lower true detection rates in the S/N regime that is best for 
          continuously sampled data.}
\label{snr_vs_fap_1}
\end{figure}

Figure~\ref{snr_vs_fap_0} shows the variation of these rates as a 
function of $\rm S/N_{\rm sp}^{\rm min}$. For the particular 
dataset we tested, FAR drops quickly near zero for $\rm S/N_{\rm sp}^{\rm min}$ 
in [5,7] and increasing number of the frequencies at highest peaks 
come into agreement with the injected signal frequencies. Unfortunately, 
the convergence interval could be shifted to other values of 
$\rm S/N_{\rm sp}^{\rm min}$ if the total time span is not covered 
uniformly, for instance, unlike in the case shown in Fig.~\ref{snr_vs_fap_0}. 
Aliasing increases the number of high peaks, thereby decreasing the 
chance of hitting the correct frequency at the highest peak. This situation 
is illustrated in Fig.~\ref{snr_vs_fap_1}, where except for the gap 
between the two datasets, the same sets and signals are used as 
above.\footnote{To be more specific, we use the same random numbers 
to generate the noise and signal parameters. However, because of the 
gap between H0 and T0, and because of the fixed position of the transit 
in T0, the same signal is shifted in time in H0 in respect to its 
original epoch in the case of continuous data sampling.} 

In spite of the higher FAR for the gapped data in the standard 
S/N regime, we caution (again) that our simple criteria for frequency 
identification lack deeper examination of the spectra. It may be 
that a more sophisticated peak statistics that also includes alias 
components would reveal somewhat better detection rates even in the 
case of gapped datasets.  

%
%################
% Figure 10
%################
%
\begin{figure}
 \vspace{0pt}
 \centering
 \includegraphics[angle=0,width=75mm]{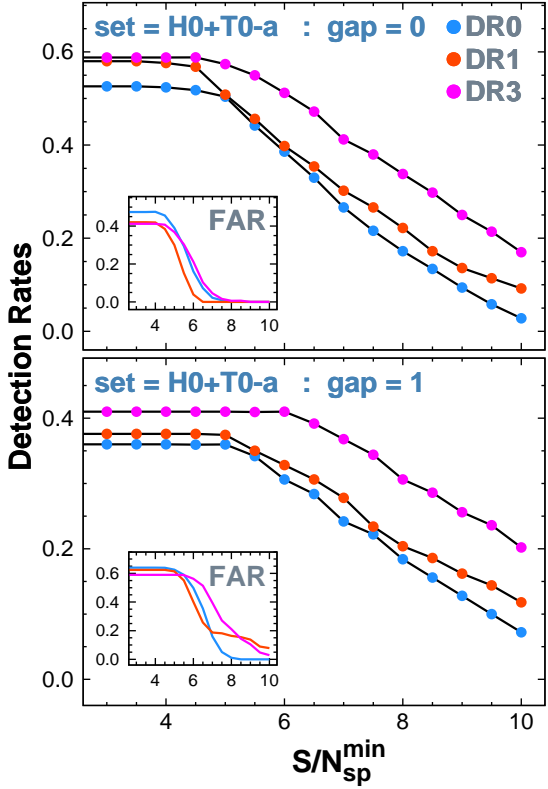}
 \caption{True detection rates for dataset H0+T0-a using various 
          signal search methods (BLS0, BLS1, and BLS3 with respective  
          detection rates of DR0, DR1, and DR3). The upper and lower 
	  panels display the results obtained with zero and $1$~yr gap 
	  between H0 and T0. The FARs used to correct the observed 
	  detection rates are displayed in the insets. For gapped data, 
	  the FAR increases considerably, leading to a decrease in the 
	  true detection rates.}
\label{dr013_2_1}
\end{figure}
%

%
%################
% Figure 11
%################
%
\begin{figure}
 \vspace{0pt}
 \centering
 \includegraphics[angle=0,width=75mm]{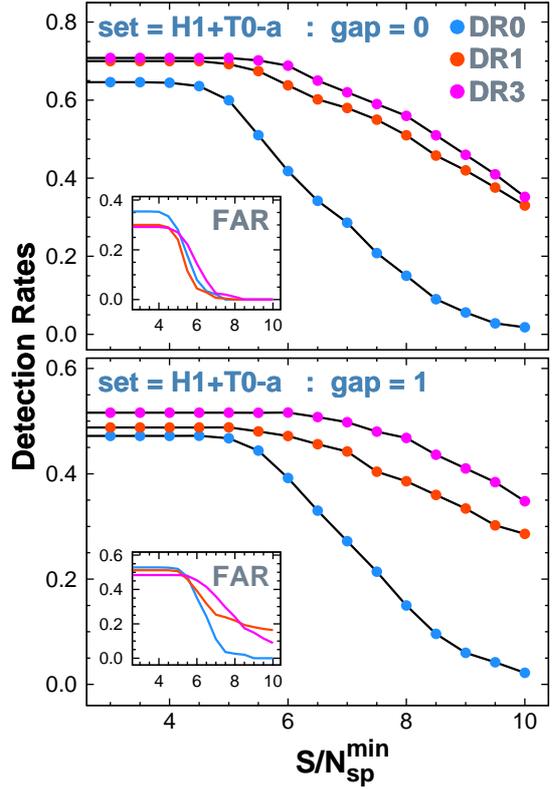}
 \caption{As in Fig.~\ref{dr013_2_1}, but for set H1+T0. We observe a 
          similar pattern as for H0+T0, with competing effects of the 
          gapped nature of H1 and the substantial increase in the size 
          of data for the same dataset.}
\label{dr013_2_2}
\end{figure}

In the following we test the detection capability of the three methods 
suggested in Sect.~\ref{bls3}. To compare the effect of gaps both 
between the primary and the secondary sets, and those in the secondary 
set alone, we used sets H0+T0 and H1+T0 with and without a $1$~yr gap 
between the primary and secondary sets. Figure~\ref{dr013_2_1} shows 
the variation in detection rates as a function of the $\rm S/N_{\rm sp}$ 
cutoff. The important common feature in both the gapped 
and non-gapped cases is the same hierarchy of the three methods. 
BLS3 outperforms the other two methods throughout the signal-dominated 
regime (e.g., for $\rm S/N_{\rm sp}^{\rm min} > 5.5$). 

The somewhat more realistic setting, with the secondary set H1 
(where the sampling is interrupted by daily and longer gaps) 
Fig.~\ref{dr013_2_2} shows that the properties mentioned above are 
retained, with some modification in the actual statistics. Although 
BLS3 still outperforms BLS1, the difference becomes less significant. 
The detection rates increase because of the larger data volume possessed 
by H1 (which apparently wins over the gaps within the dataset, which
work against the higher detection rate). Compared with the FAR displayed 
in Fig.~\ref{dr013_2_1}, we observe a lower decline, also attributed to 
the gapped nature of the secondary set. 

%
%################
% Figure 12
%################
%
\begin{figure}
 \vspace{0pt}
 \centering
 \includegraphics[angle=-90,width=75mm]{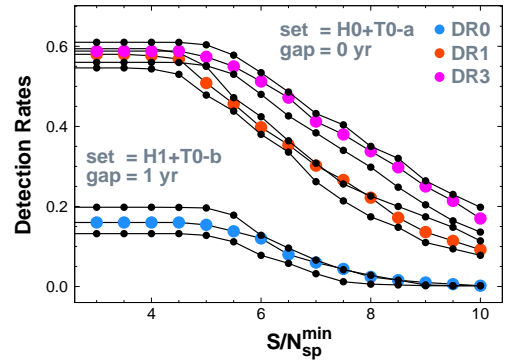}
 \caption{Testing the effect of the noise realization. Dots colored 
          other than black denote our standard random number 
          initialization. Smaller black dots show the results obtained 
          with different seeds for the random number generator. Upper 
          and lower lines for these tests for each standard simulation 
          resulted from the same pair of seeds. Labels have the same 
          meaning as in Fig.~\ref{dr013_2_1}.}
\label{dr_iseed}
\end{figure}

Because the detection rates result from random simulation, it is a 
valid concern whether the number of realizations (i.e., $500$ in our
tests) is enough for drawing reliable conclusions from 
these simulations. We note that these simulations are meant 
to cover a relatively large parameter space consisting of $P_{\rm orb}$, 
$\delta,$ and $\sigma,$ and for each given set of parameters, the 
time series generated by the particular noise realization. Therefore
we performed additional tests, with different seeds for the random 
number generator, to show how the detection rates change. For simplicity, 
we chose H0+T0 with signal type~{\em a} and zero gap. For compatibility 
with the basic tests, we used $500$ realizations. The result is displayed 
in Fig.~\ref{dr_iseed}. It is clear that there is some dependence on the 
realization, but the relative relation of the different 
search methods still remain essentially the same, that is, the order of 
preference of the different methods does not change (neither qualitatively
nor quantitatively). 

%
%################
% Figure 13
%################
%
\begin{figure}
 \vspace{0pt}
 \centering
 \includegraphics[angle=-90,width=75mm]{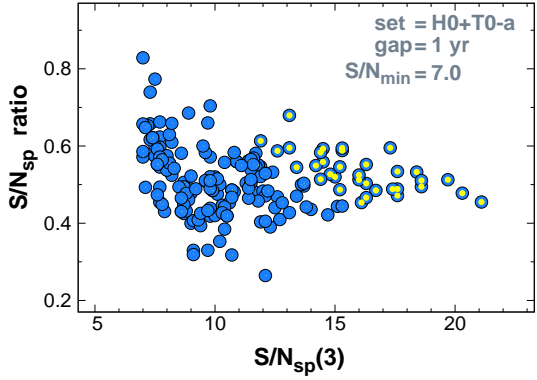}
 \caption{Comparison of the number of detections from the joint analysis 
          by BLS3 (blue points) with those resulting from the standard 
          BLS analysis of the secondary dataset H0 alone (yellow points). 
          Spectrum S/N for the analysis by BLS3 on `H0+T0' is denoted 
	  by $\rm S/N_{\rm sp}(3)$. The ordinate shows the ratio of the 
	  spectrum S/N values for `H0 alone' over `H0+T0'. Signal type~{\em a} 
	  with $500$ realizations was used.}
\label{h_vs_ht}
\end{figure}

Next we addressed the natural question whether the traces of the 
signal can be detected even in the secondary dataset without using any information 
from the primary set. If so, then what percentage of them are reliable 
detections? We chose two examples to illustrate the remarkable signal -boosting capability of the high-S/N single transit in the primary 
dataset T0.  

In Fig.~\ref{h_vs_ht} we plot the ratio of the spectrum S/N values 
for the `H0 alone' and `H0+T0' cases as a function of $\rm S/N_{\rm sp}$ 
for BLS3. In an overwhelming majority of cases, the signals 
would have remained undetected in the standard BLS analysis of the 
secondary set H0 alone. In the $500$ realizations of signal type~{\em a,} 
the BLS3 analysis discovers $37$\% of them. This can be 
compared with the $7$\% success rate from analyzing H0 alone. 

%
%################
% Figure 14
%################
%
\begin{figure}
 \vspace{0pt}
 \centering
 \includegraphics[angle=-90,width=75mm]{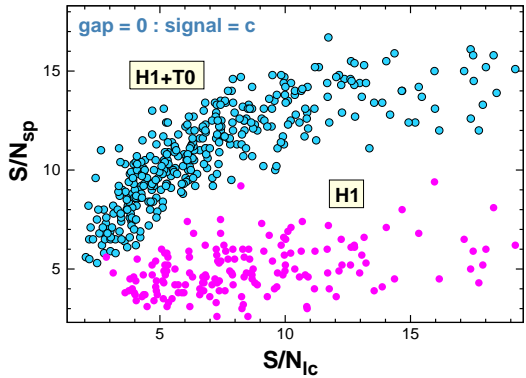}
 \caption{Signal-to-noise ratios derived from phase-folded time series 
          $\rm S/N_{\rm lc}$ vs. those of the corresponding BLS spectra,  
          $\rm S/N_{\rm sp}$, for simulations satisfying the frequency 
          criterion, condition {\em b} in Sect.~2.3. The joint 
          analysis (by using BLS3) is capable of boosting even those 
          signals that are otherwise buried deep in the noise (i.e., 
          have $\rm S/N_{\rm lc} < 3.0$ in the `H1 alone' standard BLS 
          analysis).}
\label{snr_sp_vs_lc}
\end{figure}

Yet another way of considering the detection rate increase by the joint 
analysis is to check if in any given case the strength of the signal 
implies detection, and if so, whether it is detected in the secondary 
set. Unfortunately, there is no parameter that would be based solely 
on the significance of the signal in the folded light curve and  would 
predict detectability. The reason for this is that the folded light 
curve lacks basic information on other signal components such as noise 
with semi-periodic components or real signals. Nevertheless, 
$\rm S/N_{\rm lc}$ (see Eq.~\ref{snr_lc_def}), computed from the signal 
parameters as realized in H1, might still have some relevance, and we 
tested its capability here.  

First of all, we note that we were able to perform this test in the present 
case because the signal parameters were known from the simulations. 
For real data, $\rm S/N_{\rm lc}$ can only be estimated 
probabilistically because the number of in-transit data points is not 
known because it is the function of the period (which is not known either). 

We found that the result of this test is quite similar for all 
signal types discussed in this paper. One example is shown in 
Fig.~\ref{snr_sp_vs_lc} for signal type~{\em c} (the case when most 
of the signals are strong, but so is the ambient noise). The joint 
analysis amplifies the true signal content not only for stronger 
signals of $\rm S/N_{\rm lc}>5$, but even for very faint signals 
with $\rm S/N_{\rm lc}<3$. The detections in the `H1 alone' analyses 
remain mostly in the low $\rm S/N_{\rm sp}$ regime, depending only 
mildly on $\rm S/N_{\rm lc}$. Consequently, the detection rate for 
BLS3 on H1$+$T0 is much higher, exceeding the rate of the `H1 alone' 
analysis by several factors. For instance, with $\rm S/N_{\rm min}=6.0,$ 
we find observed rates of $0.83$ and $0.09$ with respective FARs of $0.12$ 
and $0.16$.

It is interesting to examine how the behavior of the weight factor $\alpha$ 
changes when multiple transit events are observable in the TESS (T0) data. It is 
expected that the optimum weight switches back to the standard inverse-variance 
weighting that leads to the minimum variance of the combined dataset. To check 
this inference, we used the H0+T0 set with zero gap between H0 and T0. Except for 
the phase, we fixed all signal parameters, namely $P=22.3$~d, $\delta=0.002$, 
$\sigma_{\rm H0}=0.003$, and used BLS0 on $\text{ten}$ random realizations. By shifting 
the signal in phase, we can control the number of transit events occurring in 
set T0. 

\begin{figure}
 \vspace{0pt}
 \centering
 \includegraphics[angle=-90,width=75mm]{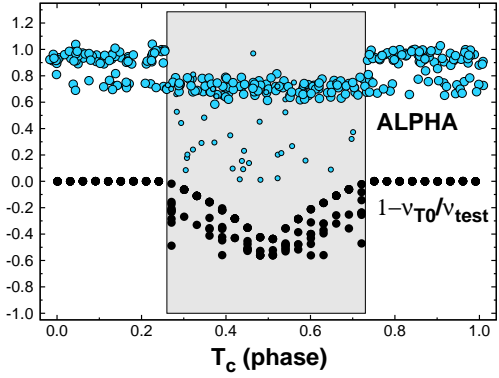}
 \caption{Dependence of the weight factor $\alpha$ of BLS3 on the phase  
          of the transit center in the case of intermediate periods, 
          when two transits may be observed in the TESS data (T0) alone. 
          The single-transit phase interval is indicated by the gray shaded 
          box. The optimum $\alpha$ values are shown by the overlapping 
          blue points (smaller dots mean no detection from the $\text{ten}$ realizations 
          we tested). We added a small noise to each point resulting from the 
          grid scan to highlight the population density. Black dots 
          indicate the relative peak frequency values for the primary dataset 
          T0, indicating detection in the two-transit regime, and lack of 
          it in the single-transit regime. See text for additional details.}
\label{scan_phase}
\end{figure}

Figure~\ref{scan_phase} shows that in the single-transit regime, the optimum 
$\alpha$ is relatively constant around the value of $0.75$. In the two-transit 
regime (i.e., outside the gray shaded box), $\alpha$ is mildly bimodal. Most 
of the values are very close to $1.0$, as expected from the inverse-variance 
weighting, predicting $\alpha=0.003^2/(0.003^2+0.0005^2)=0.97$. In some cases 
the optimum single-transit values are preferred. A closer examination of the 
run of $\rm S/N_{\rm sp}(\alpha)$ shows that this function tends to be double-humped, 
and depending on the phase, noise realization, and data distribution, the lower $\alpha$ values are preferred in rare cases even when multiple transits 
are available from the primary dataset T0. For comparison, we also show the 
relative frequency distance $1-\nu_{\rm T0}/\nu_{\rm test}$ for the peak 
frequency of the spectrum of T0. The signal is detectable 
in the two-transit regime, and as expected, cannot be recovered in the 
single-transit regime. 

%
%################
% Figure 15
%################
%

To further assess the signal detection capability of the joint 
analysis of the space- and ground-based data, we tested the dataset H2+T0, 
with H2 spanning over three years and containing some forty-five 
thousand data points (see Table~\ref{test_data}). Our test was very 
limited: a) we assumed that this dataset was placed immediately before 
the TESS data, that is, there is   no gap between them. b) We fixed the 
period to two values, $P=17.3$~d to test the short-period regime with 
transit phase allowing only a single transit in the TESS data, and 
$P=25.3$~d to test the longer period, close to the regime when 
only single transits are possible.  
c) The noise was fixed for H2 to $\sigma_{\rm H2}=0.003$ 
and for T0 to $\sigma_{\rm T0}=0.0002$. d) The weight factor was set to 
$\alpha=0.97$, based on few detailed runs for the optimization of this 
parameter. e) The transit depth was scanned in four values: 
$(0.02,0.04,0.06,0.08)$\%. f) For each transit depth, we ran $100$ random 
simulations to gain some information on the effect of noise realization. 
The transit duration was computed in the same way as in the case of 
the basic simulations discussed earlier (see Table~\ref{test_data}). 
Although these simulations are by no means meant to fully characterize 
the detection capability of H2 with T0, they give at 
least some indication on the accessible planet population in an almost 
ideal case. 

%
%================
% Table 3
%================
%
\begin{table}[!h]
  \caption{Extended dataset (H2+T0) detection rates.}
  \label{h2+t0}
  \scalebox{0.9}{
  \begin{tabular}{ccccccc}
  \hline
  \hline
  P[d]  &  $\delta$[\%] &  $\rm DR_{\rm H2+T0}$ &  FAR  & $\rm <S/N_{\rm sp}>$ &  $\rm DR_{\rm H2}$ &  FAR \\
 \hline
 17.3   &  0.08         &         1.000         & 0.030 &       10.7     &       0.810        & 0.099 \\
        &  0.06         &         0.990         & 0.141 &        9.0     &       0.350        & 0.400 \\
        &  0.04         &         0.850         & 0.612 &        7.4     &       0.180        & 0.944 \\
        &  0.02         &         0.340         & 0.971 &        6.8     &       0.190        & 1.000 \\
 25.3   &  0.08         &         0.850         & 0.176 &        7.7     &       0.350        & 0.686 \\
        &  0.06         &         0.700         & 0.586 &        6.9     &       0.270        & 0.963 \\
        &  0.04         &         0.650         & 0.908 &        6.8     &       0.310        & 1.000 \\
        &  0.02         &         0.360         & 1.000 &        6.7     &       0.330        & 1.000 \\
\hline
\end{tabular}}
\begin{flushleft}
\underline{Notes:}
{\small 
Observed detection rates for $\rm S/N_{\rm min}=6$ are shown. The true rates can 
be calculated through $\rm DR_{\rm true}=(1-{\rm FAR)}\times DR_{\rm obs}$. 
$\rm S/N_{\rm sp}$ is computed in the $\pm 0.015$d$^{-1}$ neighborhood of the 
test frequency and refers to the joint analysis. See text for additional details 
of the tests.}
\end{flushleft}
\end{table}

The detection rates and some accompanying quantities for these 
simulations are displayed in Table~\ref{h2+t0}. The lower detection rates 
for the longer period case are attributed to the lower number of in-transit 
data points ($294$ vs $587$ in the short-period simulations). Nevertheless, 
the sub-ppt regime down to $\sim 0.5$~ppt, although with substantially 
increased FARs, are accessible in both cases. While for 
higher S/Ns at $\delta=0.6$~ppt the ratio of the joint 
versus `H2 alone' detection rates is only 4 in the short period case, 
this increases to 30 at $\delta=0.4$~ppt. Shifted to larger transit 
depths, the situation is similar for the longer period case.    

\begin{figure}
 \vspace{0pt}
 \centering
 \includegraphics[angle=-90,width=75mm]{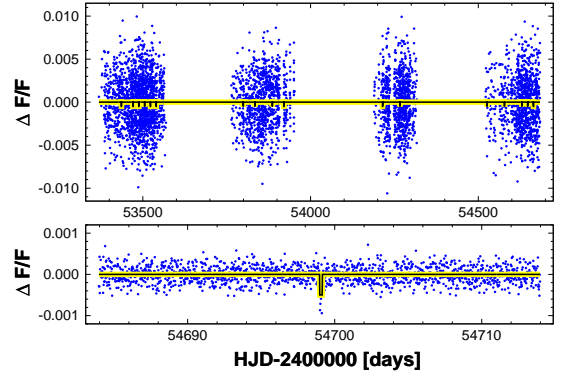}
 \caption{Simulated time series exhibiting the overall characteristics 
          of the signals generated on the time distribution of sets H2 
          and T0 (upper and lower panels, respectively). The synthetic 
          signal is shown by the black continuous line (in a yellow silhouette 
          for better visibility). The signal parameters are 
          $P=17.3$~d, $\delta=0.0005$, $t14/P=0.011$, 
          $\sigma_{\rm T0}=0.0002$, and $\sigma_{\rm H2}=0.003$ (see also 
          Table~\ref{test_data}). Only every tenth data point  
          is plotted for H2 for clarity.}
\label{h2-t0_ts}
\end{figure}

\begin{figure}
 \vspace{0pt}
 \centering
 \includegraphics[angle=-90,width=75mm]{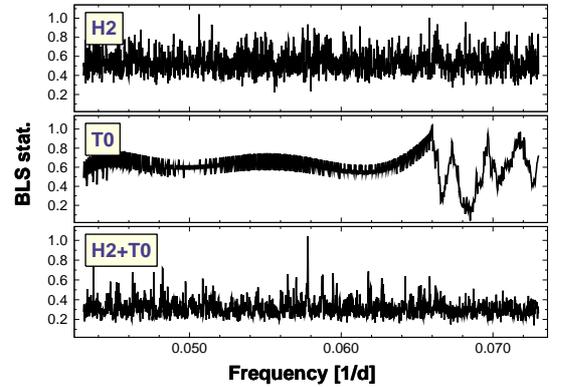}
 \caption{Frequency spectra of the time series shown in Fig.~\ref{h2-t0_ts}. 
          We use $\alpha=0.97$ in the computation of the BLS3 spectrum of 
          the combined dataset. The signal is not detected 
          in the `H2 alone' analysis (for T0 the detection is ab initio 
          excluded due to the lack of multiple transit events).}
\label{h2-t0_sp}
\end{figure}

To illustrate the signal detection power of the joint analysis in 
a `twilight zone' (i.e., at the verge of detection) case, Figs.~\ref{h2-t0_ts} 
and~\ref{h2-t0_sp} show the time series and the frequency 
spectra, respectively. For comparison, in the upper two panels of Fig.~\ref{h2-t0_sp} 
the spectra of the separate analyses of H2 and T0 are also shown.

%
%################
% Figure 16
%################
%
%

%
%################
% Figure 17
%################
%
%

%  
%%%%%%%%%%%%%%%%%%%%%%%
% SECTION 4
%%%%%%%%%%%%%%%%%%%%%%%
%
\section{Conclusions}\label{conclusions}
A combination of ground- and space-based observations is usually complimentary 
because of the different precision, wavelength, resolution, time span, etc. 
There are also ultraprecise data (i.e., those gathered by the Kepler satellite) 
that remain in the realm of space observatories. Nevertheless, even in 
these cases, ground-based data might be extremely useful when the 
signal is not hopelessly below the detection limit of the ground-based 
instruments. We investigated a case that falls in this category:   
a scenario when only a single transit is observed from space. We showed 
that an optimum combination of these and the ground-based data leads to a secure 
transit signal detection, allowing full photometric characterization of 
the system, including the period. The main steps of the suggested method 
are listed below. 

\medskip
\noindent
{\em Step0:} Determining the transit parameters $T_{\rm c}$ 
(transit center) and $t14$ (transit duration) from the space data 
$\rightarrow$ 

\noindent
{\em Step1:} Transit search with these parameters fixed and the ground-based 
data heavily weighted 
$\rightarrow$ 

\noindent
{\em Step2:} Repeating the search with free-floating transit parameters, 
a standard BLS search with fixed weights: $\alpha$ for the space- and 
$1-\alpha$ for the ground-based data  

\noindent
{\em Step3:} Inverse-variance averaging of the normalized frequency 
spectra of {\em Step1} and {\em Step2}  
$\rightarrow$ 

\noindent
{\em Step4:} Repeat {\em Step2} by scanning the weight $\alpha$ to find 
its optimum value by maximizing the S/Ns of the average 
spectra.  
\medskip
 
We found that the optimum weight on the ground-based data in {\em Step1}  
is between $0.9$ and $1.0$ in almost all cases. To save execution 
time, we can therefore fix this weight in this interval. On the other hand, 
weight $\alpha$ is a more complicated function of the actual data and 
signal settings. Nevertheless, we found that it is always greater than 
$0.5$. Therefore, this parameter should be scanned in $[0.5,1.0]$ and its 
optimum value be determined according to {\em Step4}. Depending on the grid 
on $\alpha$, all these steps lead to a multifold increase in the execution 
time. In a preliminary survey of the data, {\em Step0} and {\em Step1} 
may already be enough, because the spectrum derived in {\em Step1} 
might have sufficient S/N to identify the period. However, as discussed 
in this paper, the spectra obtained after performing full optimization 
({\em Step3}) will certainly be of higher quality, and thereby lead to 
detections of higher confidence. In crucial cases of low S/N, the full 
four-step procedure above may be the only way to determine the correct 
period.    

%
%################
% Figure 18
%################
%
\begin{figure}
 \vspace{0pt}
 \centering
 \includegraphics[angle=-90,width=75mm]{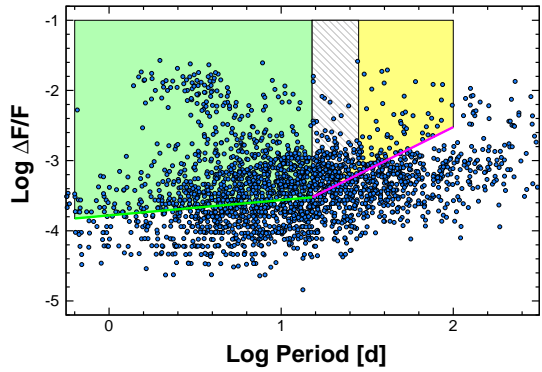}
 \caption{Accessible transiting extrasolar planets for TESS single-sector 
          observations (shaded regions) in the period -- transit depth 
          diagram. The green shaded polygon covers the region where the 
          data from TESS alone are enough for the detection. The 
          yellow shaded region shows where ground-based data are required, 
          whereas in the hatched intermediate region, a proper transit phase 
          might make the detection possible using the TESS data alone. Dots 
          denote the confirmed transiting extrasolar planets according to 
          the NASA Exoplanet Archive. See text for details on the detection 
          limits.}
\label{neptunes}
\end{figure}

The detection power of the optimum-weighted joint analysis surpasses that 
of an analysis that is based on ground-based data alone by a factor of 
$2$--$10$. Consequently, no detection in an analysis based on ground-based 
data alone does not mean that these data are not useful. Naturally, the 
generic problem of a period search in gapped data also holds in this case: 
larger gaps make the detection less likely because of the increased complexity 
of the frequency spectrum. Therefore, continuation of the ground-based 
surveys even under running precise space-based surveys is strongly preferred, 
as compared to relying only on data gathered several years ago. Naturally, 
both to decrease the noise level and to increase the duty cycle, 
a combination of various ground-based survey data is very useful.  

As indicated in Sect.~\ref{efficacy}, even under ideal circumstances, 
the combination of ground- and space-based surveys is unlikely to 
discover transits shallower than $\sim 0.03$\% (see Table~\ref{h2+t0}). 
Even so, this is a very remarkable lower limit that enables us to sample 
the Neptune -- sub-Neptune population quite deeply. Figure~\ref{neptunes} 
shows the region that is expected to be covered by the joint analysis of the 
space- and ground-based data in the orbital period -- transit depth plane. 
We used the NASA Exoplanet 
Archive\footnote{See \url{https://exoplanetarchive.ipac.caltech.edu/} 
as of 2019-02-09.} to show the currently known population of 
confirmed transiting extrasolar planets. The lower limits on 
the transit depth are highlighted in green and fuchsia. 
These limits are based on the simulations presented in this paper, 
supplemented by some additional simulations on dataset H2 of 
Table~\ref{test_data}. 

The regime of the jointly discovered planets is expected to be  
confined mostly to the relatively sparsely populated part of this 
diagram. It would be important to sample this part more effectively 
because these systems will be important in the near future, when lower 
temperature gas giants (more similar to our Neptune) will be the targets 
of atmospheric characterization.

%%%%%%%%%%%%%%%%%%%%%%%
% Acknowledgements
%%%%%%%%%%%%%%%%%%%%%%%
%
\begin{acknowledgements}
It is a pleasure to thank the referee for the constructive comments.  
Support from the National Research, Development and Innovation Office 
(grants K~129249 and NN~129075) is acknowledged. 
\end{acknowledgements}

\bibliographystyle{aa} % style aa.bst

\end{document}